# Telecom-band Chiral Light Detection through Hidden Giant Third-Order Nonlinear Circular Dichroism in Two-dimensional Halide Perovskite

*Yuki Takahashi [1], Daichi Okada [1]*, Koshi Oi[2], Kazuki Mitamura[2], Takahiko Endo[3], Yasumitsu Miyata[3], Taishi Takenobu [2] and Kenichi Yamashita[1]*

*1, Faculty of Electrical Engineering and Electronics, Kyoto Institute of Technology, Matsugasaki, Sakyo-ku, Kyoto, 606-8585, Japan*

2, *Department of Applied Physics, Graduate School of Engineering, Nagoya University, Furo-cho, Chikusa-ku, Nagoya, Aichi 464-8603, Japan*

3, *Research Center for Materials Nanoarchitectonics, National Institute for Materials Science, Tsukuba, 305-0044, Japan*

* e-mail: oka-d@kit.ac.jp

## Abstract

Chiral nonlinear optical (NLO) responses enable efficient discrimination of circularly polarized light (CPL) and are attracting increasing interest for optical and optoelectronic technologies. However, studies on NLO properties in chiral materials have largely focused on second-order NLO processes, while the role of chirality in third-order NLO processes remains poorly explored. Here, we demonstrate telecom-band CPL detection by third harmonic generation circular dichroism (THG-CD) in the chiral two-dimensional perovskite (R/S-MBACl)$_2$PbI$_4$ and uncover a giant hidden THG-CD anisotropy that is accessible *via* polarization-resolved detection. Polarization-resolved THG measurements reveal that opposite chiral NLO responses emerge in orthogonal THG polarization channels. Therefore, these chiral anisotropic contributions largely cancel each other in the total-THG signal detection, leading to underestimation of the THG-CD dissymmetry in conventional evaluations based on total-THG intensity. By separating these hidden chiral contributions, we observe exceptionally large dissymmetry factors exceeding ±1.9 and achieve selective extraction of chiral NLO responses with opposite handedness, without any structural chiral inversion. These findings highlight the importance of polarization-resolved analysis for evaluating more accurate chiral NLO responses inherent to the material and provide a promising platform for optical information processing, encryption, and anti-counterfeiting technologies at technologically relevant telecommunication wavelengths.

## Introduction

Discrimination of the handedness of chiral information, such as circularly polarized light (CPL) and spin-polarized electrons, provides an additional degree of freedom for optoelectronic systems and enables advanced information processing. Chiral materials offer an ideal platform for such functionalities, as their intrinsic structural chirality induces anisotropic interactions with chiral information carriers. [1-5] For chiral light discrimination, the most fundamental manifestation is circular dichroism (CD), which arises from differential absorption of left and right-handed CPL. In most materials, however, the magnitude of CD is typically far less than one percent, which severely limits its practical applicability. For practical applications, more efficient approaches for the discrimination of CPL are therefore required.

Among the various optoelectronic phenomena, nonlinear optical (NLO) responses and photocurrent generation are widely recognized as powerful properties for CPL discrimination. Upon irradiation or excitation with CPL, the NLO signal [6-10] or generated photocurrent [11-16] can exhibit pronounced handedness dependence, with the difference reaching several tens of percent. Such remarkable chiroptical sensitivity has fueled intense interest in exploration of novel chiral materials or system for achievement further better sensitivity.  In particular, chiral organic–inorganic hybrid materials provide a versatile platform that combines the structural tunability of organic molecules with the strong light–matter interactions of low-dimensional inorganic semiconducting frameworks. This synergy gives rise to the distinctive chirality-related phenomena and opens new opportunities for superior optoelectronic applications. [17-20]

In the field of chiral nonlinear optics, a wide variety of organic-inorganic hybrid materials have been explored, leading to remarkable advances in CPL discrimination. In some cases, handedness discrimination, quantified by dissymmetry factors approaching unity, has been achieved. [21] To date, however, the majority of studies have primarily focused on second-order NLO processes, in which chirality-induced symmetry breaking plays a key roll in nonlinear chiroptical responses that can be probed through discrimination or generation of CPL, such as second harmonic generation circular dichroism (SHG-CD) [6-10] and circularly polarized second harmonic generation (CP-SHG). [22-25] In contrast, the contribution of chirality to third-order NLO responses has not yet been systematically and extensively investigated, despite the fact that organic–inorganic metal halides are known to exhibit strong third-order nonlinear optical responses due to their pronounced excitonic nature originating from low-dimensional structures. [26-31] So far, studies on chiral third-order NLO responses have been largely limited to metamaterial-based systems, including metallic helicoidal colloids exhibiting circular-polarization-dependent hyper-Rayleigh or Mie scattering [32,33], as well as third harmonic generation (THG) in chiral metamaterials [34-40]. However, investigations of the role of chirality in third-order NLO responses in organic materials and organic–inorganic hybrid materials remain essentially unexplored.

In this study, we investigate THG circular dichroism (THG-CD) in the chiral two-dimensional (2D) halide perovskite $(R/S\text{-}MBACl)_2PbI_4$. The material exhibit pronounced handedness-dependent THG responses, enabling efficient discrimination of CPL and wavelength-converted detection of telecom-band information into the visible region, which is not achievable with conventional reported SHG-CD process. Polarization-resolved measurements further reveal that the orthogonally polarized THG components possess oppositely signed THG-CD signals, which largely cancel out under conventional unpolarized total-THG detection. By separating these components by polarizer, a giant hidden chiral NLO response with dissymmetry factors exceeding $g_{\text{THG-CD}} > |1.9|$ is uncovered. Such a unique chiral response is also exploited to demonstrate a novel chiral anti-counterfeiting system. These findings establish THG-CD as a powerful platform for chiral information processing and reveal previously hidden chiral functionalities for advanced optical technologies.

## Results and discussion

Chiral two-dimensional (2D) halide perovskites, $(R/S\text{-}MBACl)_2PbI_4$, were used in this study (Figure S1). This material has been extensively investigated previously, from fundamental chiroptical properties to chiral SHG, where pronounced anisotropic SHG-CD responses have been reported [22, 41,42]. The NLO responses of single crystal (R/S-SC) and thin-film (R/S-TF) of $(R/S\text{-}MBACl)_2PbI_4$ were examined (See Materials and methods), with the primary focus placed on the R/S-SC samples. Owing to their strong excitonic character, the R/S-TF exhibit intense and spectrally sharp excitonic absorption (Figure S2). Such an excitonic character is beneficial for third-order NLO response.

THG measurement was conducted using a home-built optical setup (See Materials and methods, Figure S3). The samples were irradiated using a wavelength-tunable picosecond laser source, and the generated harmonic signals were collected using a spectrometer equipped with a CCD. The crystal was in-plane rotated and positioned so that the incident linear polarization was aligned with the crystalline axis corresponding to the maximum THG intensity. Under irradiations at 1500 nm, the R/S-SC exhibited clear frequency up-conversion into the visible color (Figure 1a). By scanning the incident laser wavelength from 1450 nm to 1650 nm, clear sharp peaks corresponding to the third-harmonic signals were observed at each incident wavelength (Figure 1b). In addition, the signal intensity exhibited a cubic dependence on the incident laser power (Figure 1c). These data provide clear evidence of THG. The THG efficiency reached its maximum at 510 nm (Figure 1b). Comparison with the absorption spectrum of the measured SC revealed that this wavelength is identical with the absorption maximum, clearly demonstrating the enhancement of THG induced by the excitonic resonance effect (Figure 1b).

The THG conversion efficiency of $(R/S\text{-}MBACl)_2PbI_4$ was investigated by comparing with CVD-grown monolayer $MoS_2$.[43] The effective NLO susceptibility of $MoS_2$ has been reported previously, therefore, it serves as a useful reference material. Because of the substantial difference in thickness between the bulk crystal and the monolayer material, a direct comparison between them cannot be considered entirely fair. Therefore, R-TF samples were used for an appropriate comparison. By comparing the THG responses of the 110 nm and 190 nm R-TFs with that of monolayer $MoS_2$ (Figure 1d), the effective third-order NLO susceptibility at $\lambda_{THG}$= 500 nm was evaluated. After considering the absorption coefficient, phase mismatch, refractive index and sample thickness, [29, 44-46] the effective NLO susceptibility of the R-TFs was estimated to be approximately 0.04 times that of monolayer $MoS_2$ (Supporting information S3). The effective NLO susceptibility of monolayer $MoS_2$ at 1500 nm has been estimated to be $10.4 \times 10^{-28}\, m^2/V^2$ [45], yielding an effective susceptibility of $4.2 \times 10^{-29}\, m^2/V^2$ for $(R/S\text{-}MBACl)_2PbI_4$. Although the extracted effective susceptibility is lower than that of monolayer $MoS_2$, its large macroscopic interaction length enables significantly stronger THG generation in the bulk crystalline state. For example, 6.5 μm thick R-SC crystal exhibits a THG intensity several thousand times greater than that of monolayer $MoS_2$ (Figure 2e).

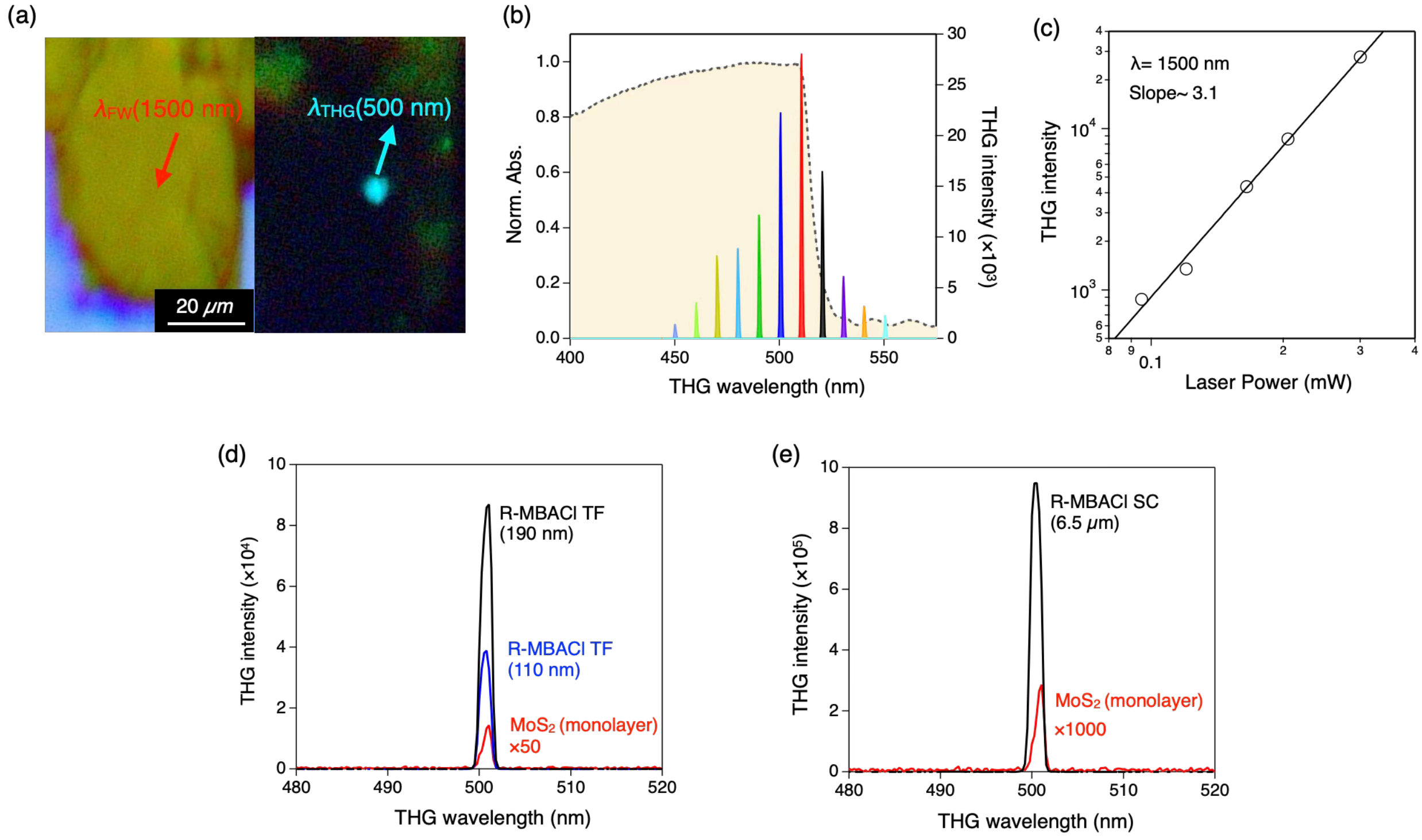

**Figure 1.** (a) Optical microscopy image of the crystal under 1500 nm laser irradiation, showing green emission at 500 nm generated *via* THG. (b) THG spectra measured under incident laser wavelengths ranging from 1450 to 1650 nm at a constant laser power of 300 $\mu$W and the normalized absorption spectrum of measured SC. (c) THG intensity as a function of incident laser power at $\lambda_{THG}$ = 500 nm. The solid line represents a power-law fit, showing the third-order nonlinear optical nature. (d, e) Comparison of THG spectra at $\lambda_{THG}$ = 500 nm between R-TF samples (110 and 190 nm) and $MoS_2$ (d), and between the R-SC sample (6.5 $\mu$m) and $MoS_2$.

Next, anisotropic response to incident circular polarization was investigated. By setting the quarter waveplate (QWP) at ±45° with respect to the polarization direction, we can choose a L- or R-CP handedness of fundamental light. Figure 2a shows the normalized THG spectra under CPL irradiation in the incident wavelength range of 1440–1560 nm, showing a clear anisotropy in the THG intensity. Figure 2b plots the wavelength dependences of the THG dissymmetry factor ($g$-factor), which is defined as $g_{\mathrm{THG-CD}} = 2\,(I_L - I_R)/(I_L + I_R)$, where $I_L$ and $I_R$ refers to the output THG intensity for the fundamental L- or R-CPL, respectively. In both R- and S-SC crystals, clear chiral light discrimination in the third order NLO response was observed throughout the measured NIR range. Although variations in the g-factor were observed among samples, these differences are likely attributable to variations in crystal quality and thickness. The power dependence of the THG intensity under CPL irradiation was also investigated. For both L- and R-CPL, the slopes of the power dependence were nearly identical, indicating that the $g$-factors is almost independent of incident laser power (Figure 2c). Figure 2d summarizes previously reported SHG-CD $g$-factors measured at fundamental wavelengths longer than 1000 nm [47-52] and compares them with the highest $g$-factors obtained for R/S-SC in Figure 2b. Among them, the longest reported CPL detection wavelength was 1200 nm in spiral [(R/S)-MBA]$_4$Bi$_2$Br$_{10}$ microcrystals, which exhibited a $g_{\mathrm{THG-CD}}$ of 0.58.[48] In contrast, the present chiral 2D halide crystals enable CPL discrimination at much longer wavelengths extending into the optical communication band, while converting these long-wavelength photons into the visible region through the THG process.

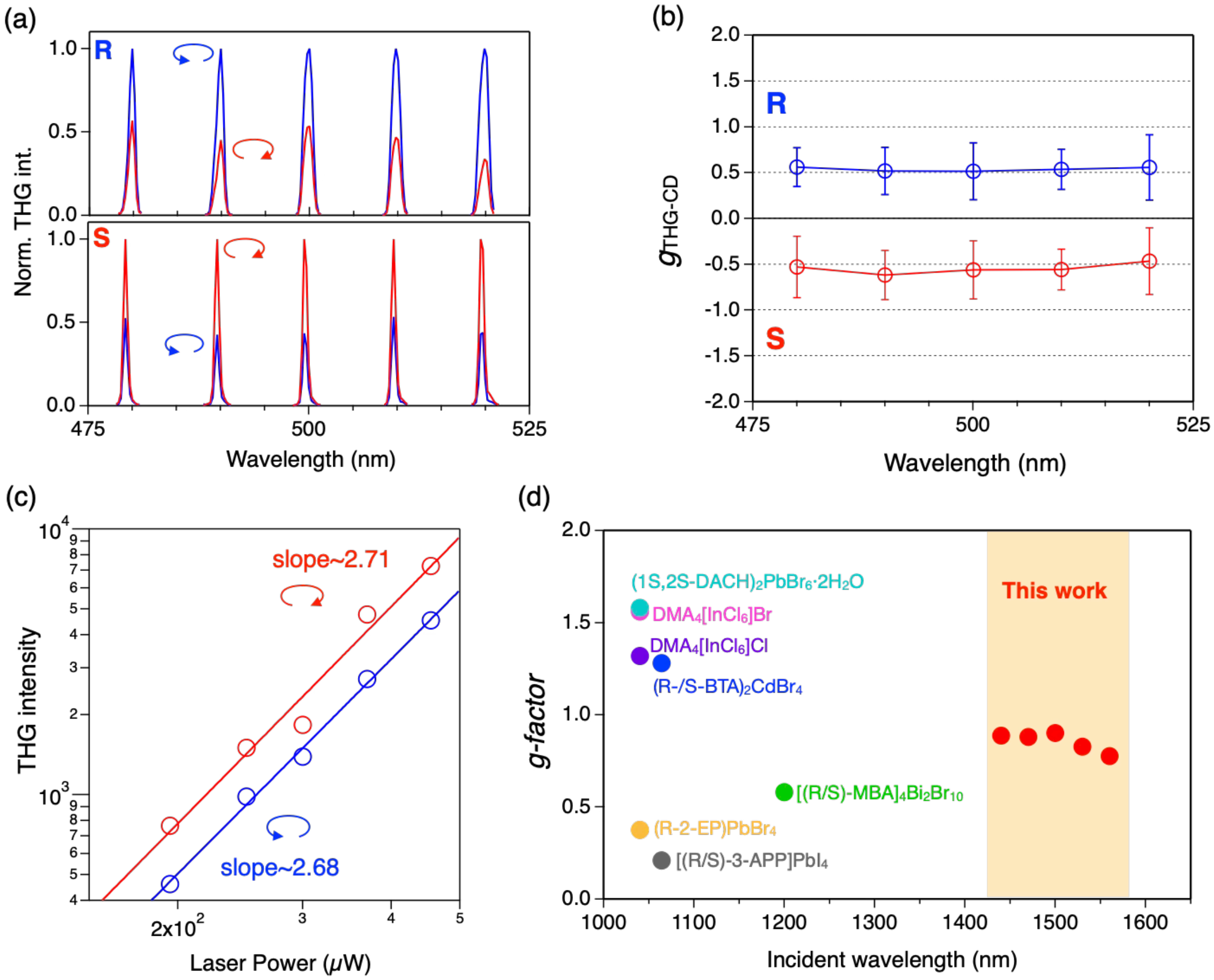

**Figure 2.** (a) Normalized THG spectra under right- and left-circularly polarized irradiation for the R- and S-SC samples, showing anisotropic handedness-dependent responses. (b) Wavelength dependence of $g_{THG\text{-}CD}$ for the R- and S-SC samples. Data represent the average $g_{THG\text{-}CD}$ values obtained from four independently measured SC sample; error bars show the standard deviation. (c) THG intensity as a function of incident laser power under RCP and LCP irradiation at $\lambda_{THG}$ = 500 nm for the S-SC sample, confirming the third-order nonlinear optical response. (d) Comparison of the $g_{THG\text{-}CD}$ values obtained in this work with previously reported $g_{SHG\text{-}CD}$ values measured at incident wavelengths longer than 1000 nm. [47-52]

To further investigate the detailed chiral response of our samples, QWP rotation analysis of the output THG was performed. In this measurement, the incident polarization state was continuously changed by rotating the QWP, while the generated THG signal was resolved into its *x*- and *y*-polarized components. The coordinate system in this study is illustrated in Figure S3. Figures 3a and 3d show the THG intensity as a function of the QWP angle for the total THG signal detected without a polarizer. Figures 3b and 3e show the corresponding x- and y-polarized THG components, and their normalized profiles are presented in Figures 3c and 3f. The conventional $g_{\mathrm{THG-CD}}$ values, defined by the total THG intensity, were estimated to be 0.70 and -0.80 for the R- and S-SC samples, respectively (Figure 3a, d). However, interestingly, when the THG signal is decomposed into *x*- and *y*-polarized components, while the *x*-polarization is dominant for total THG, chiral response of each polarization component is found to be clearly reversed (Figure 3b, c, e, f). For the R-SC (S-SC) sample, the *x*-polarized THG component exhibits a pronounced minimum, approaching nearly zero intensity, at QWP angles corresponding to R-CPL (L-CPL), while the *y*-polarized component almost reaches to its maximum under the same CPL direction. These results also indicate that the handedness of incident CPL is encoded into the orthogonal linear polarization of THG (Figure S4). For example, in case of R-SC, incident R-CPL generates THG with an *x*-polarization purity of up to 99 %, whereas L-CPL excitation predominantly produces y-polarized THG with a polarization purity of 77 % (Figure S4). Therefore, exceptionally large dissymmetry factors are obtained in such polarization-resolved condition. For the R-SC sample, $g_{x,\mathrm{THG-CD}} = -1.66$ and $g_{y,\mathrm{THG-CD}} = 1.95$ were obtained for the *x*- and *y*-polarized components, respectively. In contrast, the signs were completely reversed for the S-SC sample, yielding $g_{x,\mathrm{THG-CD}} = 1.77$ and $g_{y,\mathrm{THG-CD}} = -1.40$. These findings reveal that a giant chiral response is hidden within the polarization-resolved THG components. Since the chiral contributions of the *x*- and *y*-polarized THG signals exhibit opposite signs, they compensate each other in the total signal. Thus, conventional *g*-factor analyses based on the total NLO signal underestimate the intrinsic chiral response.

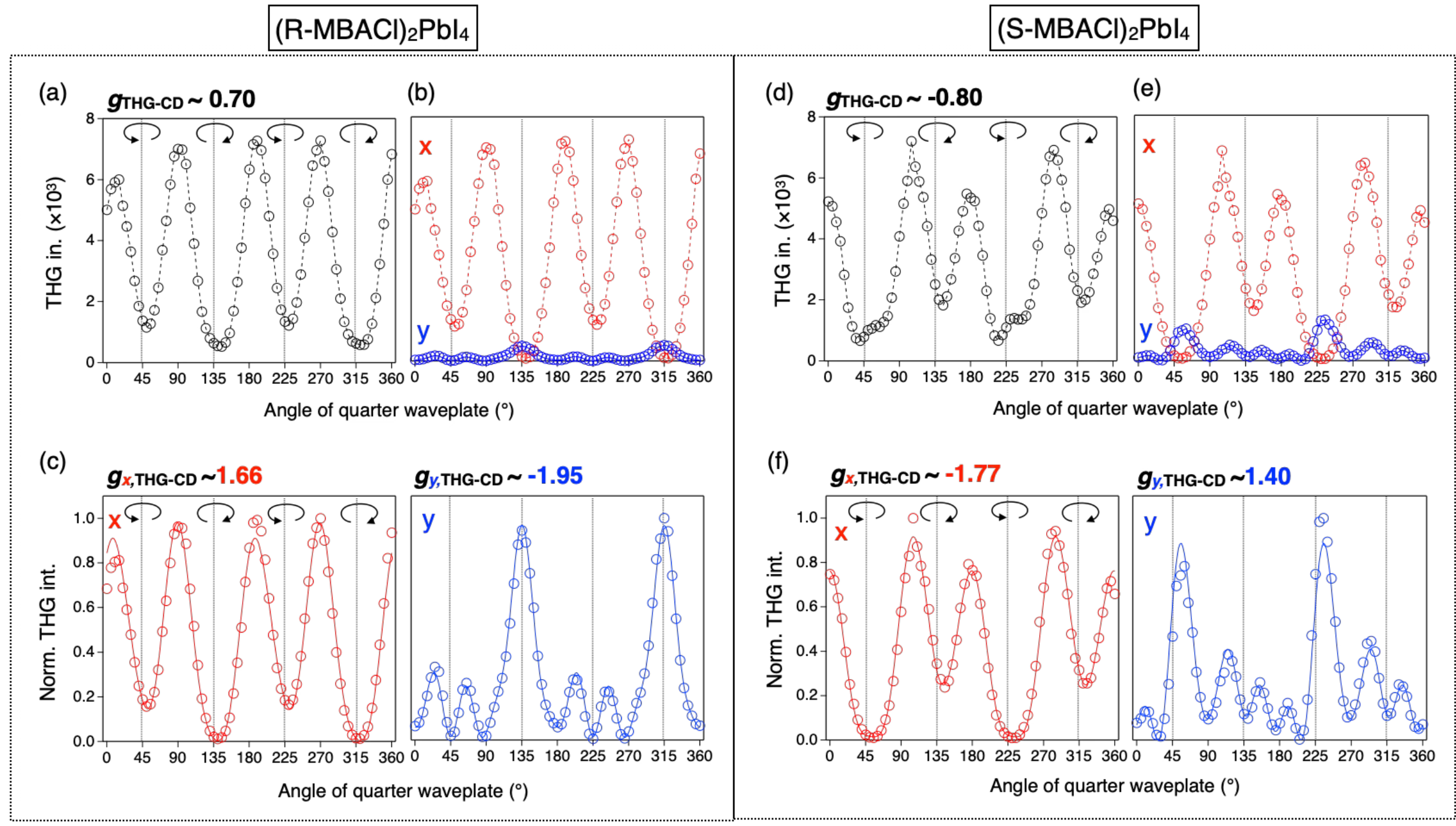

**Figure 3.** (a, d) Total-THG intensity detected without polarizer as function of QWP rotation angle for R- (a) and S-SC (d) sample. (b, e) x- (red) and y-polarized (blue) THG intensity profile as function of QWP rotation angle for R- (b) and S-SC (e) sample. (c, f) Normalized THG intensity profiles for the *x*- and *y*-polarized components obtained from Figure 3b and 3e. The solid line shows fitting curve evaluated by Eq. 4. The obtained fitting value is expressed in Table S1.

The (R/S-MBACl)$_2$PbI$_4$ SC belong to triclinic *P*1 structure [41], which possesses the lowest crystal symmetry. Owing to this low-symmetry crystal structure, all components of the third-order nonlinear susceptibility tensor remain nonzero and independent. Under the present experimental configuration employing normal incidence and normal detection, the nonlinear polarization can be expressed using the susceptibility tensor components as follows (Supporting Information S4):[53]

$$\begin{pmatrix} P_X^{(3\omega)} \\ P_Y^{(3\omega)} \end{pmatrix} = \frac{\epsilon_0}{4} \begin{pmatrix} \chi_{11} & \chi_{12} & \chi_{18} & \chi_{19} \\ \chi_{21} & \chi_{22} & \chi_{28} & \chi_{29} \end{pmatrix} \begin{pmatrix} E_X^3 \\ E_Y^3 \\ 3E_X E_Y^2 \\ 3E_X^2 E_Y \end{pmatrix} \quad (1)$$

For CPL dependent NLO response, considering the relation $E_x = \pm iE_y$, the asymmetric response to opposite circular polarizations originates from the $E_y^3$ and $E_x^2 E_y$ terms. Accordingly, the tensor components $\chi_{12}$, $\chi_{19}$, $\chi_{22}$, and $\chi_{29}$ are responsible for generating the circularly anisotropic response. And THG intensity difference between CPL handedness ($\Delta I(3\omega)_{THG-CD}$) can be expressed as:

$$\Delta I(3\omega)_{THG-CD} \propto 4Im[(\chi_{11} - 3\chi_{18})(\chi_{12} - 3\chi_{19})^*] \text{ (}x\text{-polarization)} \quad (2)$$

$$\Delta I(3\omega)_{THG-CD} \propto 4Im[(\chi_{21} - 3\chi_{28})(\chi_{22} - 3\chi_{29})^*] \text{ (}y\text{-polarization)} \quad (3)$$

These expressions indicate that the chiral response originates from the phase difference between tensor components exhibiting symmetric responses to CPL ($\chi_{11}$, $\chi_{18}$, $\chi_{21}$, and $\chi_{28}$), and those responsible for asymmetric responses ($\chi_{12}$, $\chi_{19}$, $\chi_{22}$, and $\chi_{29}$). Importantly, in the $P1$ crystal structure, the *x*- and *y*-polarized THG components are governed by different sets of independent tensor elements. Therefore, it is not surprising that the chiral response differs between the two polarization channels. Their sign and magnitude of the chiral response are determined by the amplitudes and relative phases of the contributing nonlinear susceptibility tensor elements. Accordingly, the opposite signs of the chiral response obtained for the *x*- and *y*-polarized THG components imply a reversal of the relative phase relationships among the effective tensor contributions in the two-polarization state. The angular dependence profiles of the obtained THG intensities were fitted as a function of the waveplate rotation angle $\phi$ using the Fourier series of harmonic functions (Figure 3 c, f, see Supporting Information S4):

$$I(3\omega) = A + B\cos 2\varphi + C\cos 4\varphi + D\cos 6\varphi + E\cos 8\varphi + F\cos 10\varphi + G\cos 12\varphi + H\sin 2\varphi + I\sin 4\varphi + J\sin 6\varphi + K\sin 8\varphi + L\sin 10\varphi + M\sin 12\varphi \quad (4)$$

The obtained fitting coefficients are summarized in Table S1. The sign reversal associated with chirality inversion is predominantly observed in the coefficients of the sine terms, suggesting that the chiral contribution is mainly encoded in the odd components of the angular response.

Considering the opposite chiral responses of the $x$- and $y$-polarized THG components and their mutual cancellation in the total THG signal, the absence of an observable $g_{\mathrm{THG-CD}}$ does not necessarily indicate the absence of chiral NLO activity. Rather, polarization-resolved THG measurements provide access to hidden chiral responses that are masked in the total THG intensity. To verify this hypothesis, polarization-resolved THG measurements were performed on R-SC samples showing nearly zero $g_{\mathrm{THG-CD}}$. As shown in Figure 4a, measured SC showed almost negligible THG-CD response ( $g_{\mathrm{THG-CD}} = 0.04$ ). However, polarization-resolved analysis revealed distinct chiral responses with opposite signs in the $x$- and $y$-polarized components (Figure 4a, b). Remarkably, the obtained $g$-factors reached $g_{x,THG-CD} = 1.93$ and $g_{y,THG-CD} = -1.92$, demonstrating nearly perfect discrimination of the incident CPL for each polarization component. These results highlight the importance of polarization-resolved measurements, which enable the extraction of chiral responses that may remain hidden in conventional THG-CD measurements. At the same time, the present system offers a unique chiral functionality in which the sign of a giant nonlinear circular dichroism can be switched simply by selecting the output polarization with a polarizer without changing the chiral enantiomeric structure (Figure 4e). In addition, since the emitted THG exhibits nearly pure linear polarization, reaching up to 97% *x*- and *y*-polarized THG under L-CPL and R-CPL excitation, respectively, the input CPL information can be directly retrieved from the polarization state of the output THG (Figure S5). Such a polarization resolved giant *g*-factor was also confirmed in the S-SC sample exhibiting a weak $g_{\mathrm{THG-CD}}$ response, as well as across the incident wavelength range of 1440–1560 nm (Figure S6 and S7a).

A detailed third-order tensor determination is challenging because considering the complex nature of the third-order nonlinear susceptibility make chiral THG analysis highly complicated. Nevertheless, the polarization resolved THG-CD responses showing the almost maximum *g*-factor of $|g_{\mathrm{THG-CD}}| = 2$, allowing us to know several important conditions on the relative amplitudes and phases of the contributing tensor elements. According to Equations (2) and (3), the maximum value of $g_{\mathrm{THG-CD}}$ is achieved when the symmetric and chiral asymmetric tensor contributions possess equal amplitudes, i.e., $|\chi_{11} - 3\chi_{18}| = |\chi_{12} - 3\chi_{19}|$ and $|\chi_{21} - 3\chi_{28}| = |\chi_{22} - 3\chi_{29}|$, together with a relative phase difference of $\Delta\Phi = \pm\pi/2$ (Supporting information S4). Therefore, the giant THG-CD responses observed here suggest that these interference conditions are closely satisfied in the present crystal sample, which is further supported by the observed polarization conversion from incident CPL to orthogonally polarized THG (Supporting information S4). This giant anisotropic response was observed only under resonant conditions near the excitonic transition, suggesting that excitonic absorption play an important role in establishing the amplitude balance and phase relationship required for the observed giant THG-CD response in this crystal (Figure S7b).

In addition, we investigated the THG responses of racemic (Rac)-SC samples and an achiral reference compound, $(PEA)_2PbI_4$, which crystallizes in the *P*1 space group. [54-55] In both systems, the *g*-factor remained nearly zero or very weak not only in the total THG intensity but also under polarization-resolved conditions (Figures 4c, d and S8). Furthermore, the THG activity itself under CPL irradiation was weaker than that observed in the chiral SC samples. These results indicate that chirality contributes not only to the generation of anisotropic chiral THG responses but also to the enhancement of light-matter interactions with CPL. Such a clear distinction between chiral and racemic (achiral) crystals, particularly in the presence or absence of giant hidden chiral responses, opens up a new design concept for NLO anti-counterfeiting technologies (Figure 4e). In the original state, the anti-counterfeiting tag simply acts as an infrared-to-visible wavelength conversion material and exhibits equivalent responses to opposite CPL, making the hidden information inaccessible. However, for authentic tags, polarization-resolved detection reveals hidden chiral responses that become observable only after introducing an analyzer, whereas counterfeit tags remain optically inactive toward CPL even under identical polarization-resolved conditions. Such selective emergence of chiral responses enables reliable authentication based on hidden NLO signatures (Figure 4e). Moreover, the polarization-dependent inversion of the chiral NLO response provides an additional degree of freedom for authentication. In addition to the presence or absence of hidden chiral responses, the sign reversal of the chiral NLO state induced by switching the output polarization can serve as an extra authentication parameter. Such multiplexed optical information is expected to further enhance the security level and anti-counterfeiting performance of the proposed tag system (Figure 4e).

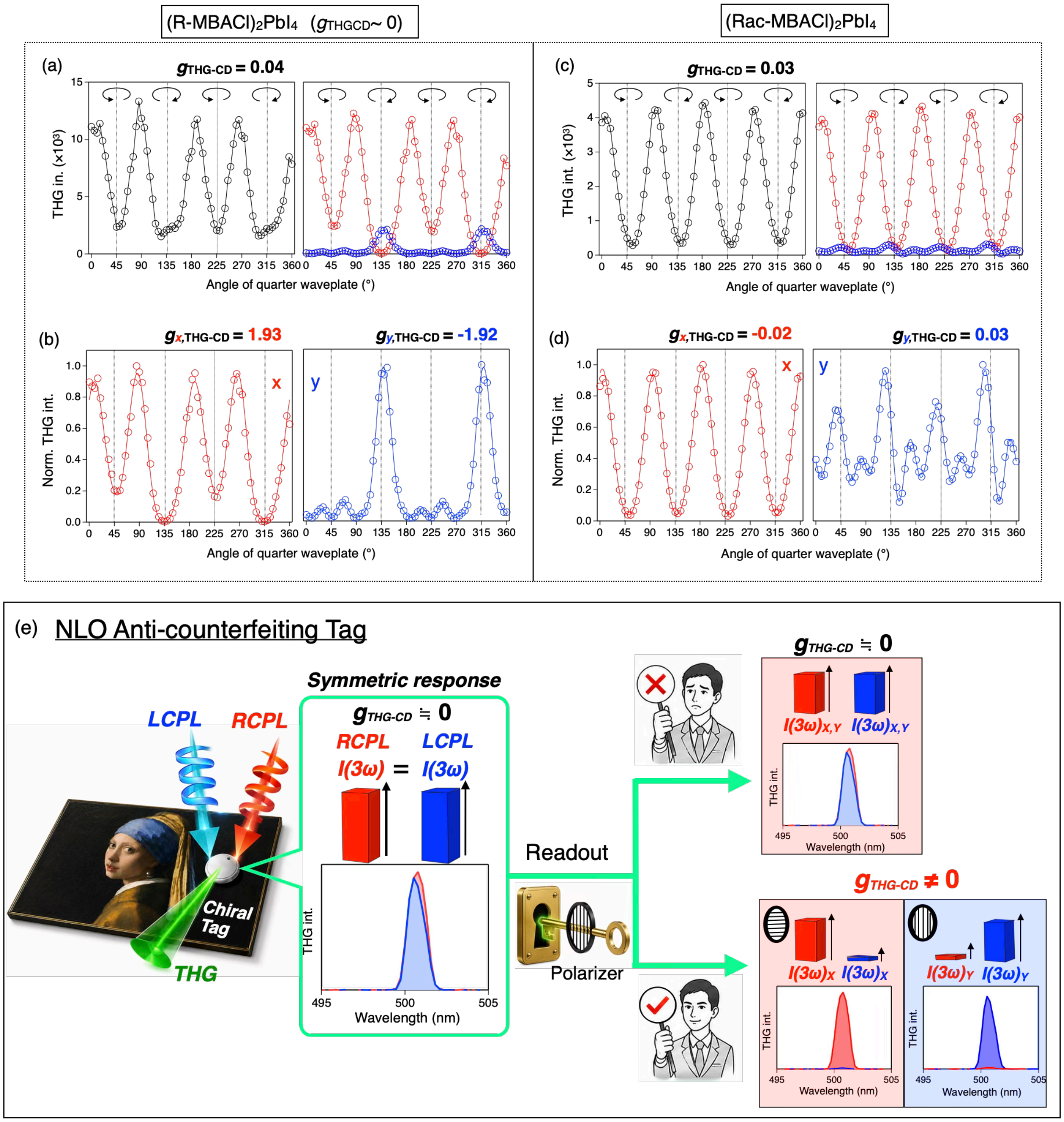

**Figure 4.** (a, c) THG intensity as function of QWP rotation angle for R- (a) and Rac-SC (b) sample. The left panel showing black dot is total THG intensity profile and right panel show x-red dot) and y-polarization (blue dot) resolved THG profile data. R-SC data show a case in which the overall g-factor is nearly negligible, whereas giant chiral responses emerge upon polarization-resolved analysis. (b, d) Normalized THG intensity profiles for the x- and y-polarized components obtained from the right panels of Figure 4a and c. The solid line shows fitting curve evaluated by Eq. 4. (e) The schematic representation of concept for potential application to anti-counterfeiting systems. The hidden information can only be accessible through polarizer.

**Conclusion**

In summary, we demonstrated efficient chiral third-order NLO responses in chiral 2D halide perovskites, arising from the combination of strong excitonic effects and pronounced chiral light-matter interactions. These characteristics enable efficient CPL discrimination through THG process at technologically important optical communication wavelengths. Beyond establishing chiral 2D perovskite as a promising platform for THG-based CPL discrimination, our study uncovers hidden chiral optical effect, whereby oppositely signed chiral responses residing in orthogonally polarized THG channels are largely concealed through cancellation in the total THG signal. By selectively accessing polarization resolved channels, giant chiral response appeared. These finding highlights that the apparent weakness of chiroptical response does not necessarily reflect the intrinsic strength of the underlying chiroptical interactions, providing a valuable evaluation framework for uncovering hidden and enhanced chiral optical responses in future material systems. Through efficient CPL discrimination beyond the spectral range of SHG and the introduction of an additional degree of freedom via polarization encoding, this THG-based platform broadens the design space of chiral nonlinear optics and paves the way for future applications in anti-counterfeiting, secure optical communications, and advanced information processing.

## Materials and methods

### Materials

Unless otherwise noted, all reagents and solvents were used as received. (R)-4-Chloro-α-methylbenzylamine (R-MBACl) is purchased from Thermo scientific. (S)-4-Chloro-α-methylbenzylamine (S-MBACl) is purchased from Apollo scientific. 57% w/w HI aqueous solution and 2-phenylehtylamine (PEA) were purchased from TCI. 50% $H_3PO_2$ aqueous solution is purchased from Sigma-aldrich. PbO was purchased from Nacalai tesque.

### Sample preparation of Single crystal and thin-film

For the synthesis of $(R/S\text{-}MBACl)_2PbI_4$ or $(PEA)_2PbI_4$ single crystals, 400 mg of PbO powder and 400 μL of R/S-MBACl or PEA were dissolved in 15 mL HI solution with 1 mL of $H_3PO_2$ as a stabilizer at 150°C until all solid was completely dissolved and a transparent yellow solution was formed. The solution was then slowly cooled to room temperature (25°C) over 48 hours, during which single crystals of 2D-OIHPs gradually precipitated. The obtained crystals were filtered, washed with toluene, and dried under vacuum for more than 48 hours. In case of synthesis for $(racemic\text{-}MBACl)_2PbI_4$, R and S-MBACl were mixed in a 1:1 ratio (200 μL: 200 μL).

The $(R/S\text{-}MBACl)_2PbI_4$ thin film was prepared by spin-coating onto a quartz glass substrate The synthesized single crystals were dissolved in DMF at a concentration and the thin film was obtained by spin-coating the solution for 60 seconds, followed by annealing at 80°C for 15 minutes. To encapsulate the sample, 100 nm PMMA film was also spin-coated on top of the 2D-OIHPs film.

The $MoS_2$ monolayer used as the THG reference was synthesized by exfoliation and CVD growth. The CVD-grown $MoS_2$ monolayer was fabricated as described in Ref. [43].

**Optical setup for THG measurement**

THG was measured by home-made optical setup as shown in Figure S2. A pico-second pulsed laser system (EKSPLA: PT403, 410 ~2300 nm) with a pulse width ~15 *ps* and a repetition rate of 1 kHz was used for fundamental incident light source. Incident laser polarization was fixed *x*-polarization by polarizer and intensity is modulated by tunable ND filter. The laser beam was focused to a spot size of ~20 μm using a plano-convex lens. The THG signal was collected by an EMCCD camera (Andor Newton) through a spectrometer (Andor Kymera 193i). For the quarter waveplate (QWP) rotational measurements, a QWP was placed before the sample to generate circularly polarized light, and an additional polarizer was set before the spectrometer to analyze the polarization state of the output THG signal. Before the THG measurements, the sample was rotated so that its crystallographic axis was aligned with the incident x-polarization, corresponding to the orientation that maximized the THG intensity.

**Data availability**

All data supporting the findings of this study are available from the corresponding author upon reasonable request.

**Authors contribution**

D.O. conceived the project, designed the experiments, and developed the overall research concept. Y.T. performed all experiments and data analyses. K.O., K.M., T.E., Y.M., and T.T. synthesized and prepared $MoS_2$ samples. D.O. and K.Y. supervised the research. D.O. wrote the manuscript, and K.Y. reviewed and edited the manuscript. All authors reviewed and approved the final version of the manuscript.

**Funding Sources**

Grants-in-Aid for Scientific Research (JP22H00215, JP22H04957, JP 23H01942, JP24KK0084, JP24H00044, JP25H00741, JP25K22242, JP25H00835, JP26K01475) from the Japan Society for the Promotion of Science (JSPS) and JST CREST (JPMJCR20T4, JPMJCR23A4,

JPMJCR23O3) and JST FOREST (JPMJFR213X) from Japan Science and Technology Agency (JST).

**Competing interests**

The authors declare no competing financial interests.

**ACKNOWLEDGMENT**

This work was partly supported by Grants-in-Aid for Scientific Research (JP22H00215, JP22H04957, JP 23H01942, JP24KK0084, JP24H00044, JP25H00741, JP25K22242, JP25H00835, JP26K01475) from the Japan Society for the Promotion of Science (JSPS) and JST CREST (JPMJCR20T4, JPMJCR23A4, JPMJCR23O3) and JST FOREST (JPMJFR213X) from Japan Science and Technology Agency (JST).

# Supporting Information

## *Telecom-band Chiral Light Detection through Hidden Giant Third-Order Nonlinear Circular Dichroism in Two-dimensional Halide Perovskite*

*Yuki Takahashi [1], Daichi Okada [1]*, Koshi Oi[2], Kazuki Mitamura[2], Takahiko Endo[3], Yasumitsu Miyata[3], Taishi Takenobu [2]and Kenichi Yamashita[1]*

[1,] *Faculty of Electrical Engineering and Electronics, Kyoto Institute of Technology, Matsugasaki, Sakyo-ku, Kyoto, 606-8585, Japan*

[2,] *Department of Applied Physics, Graduate School of Engineering, Nagoya University, Furo-cho, Chikusa-ku, Nagoya, Aichi 464-8603, Japan*

[3,] *Research Center for Materials Nanoarchitectonics, National Institute for Materials Science, Tsukuba, 305-0044, Japan*

* To whom correspondence should be addressed.
E-mail: oka-d@kit.ac.jp

## 1. Sample preparation (Single crystal, thin-film)

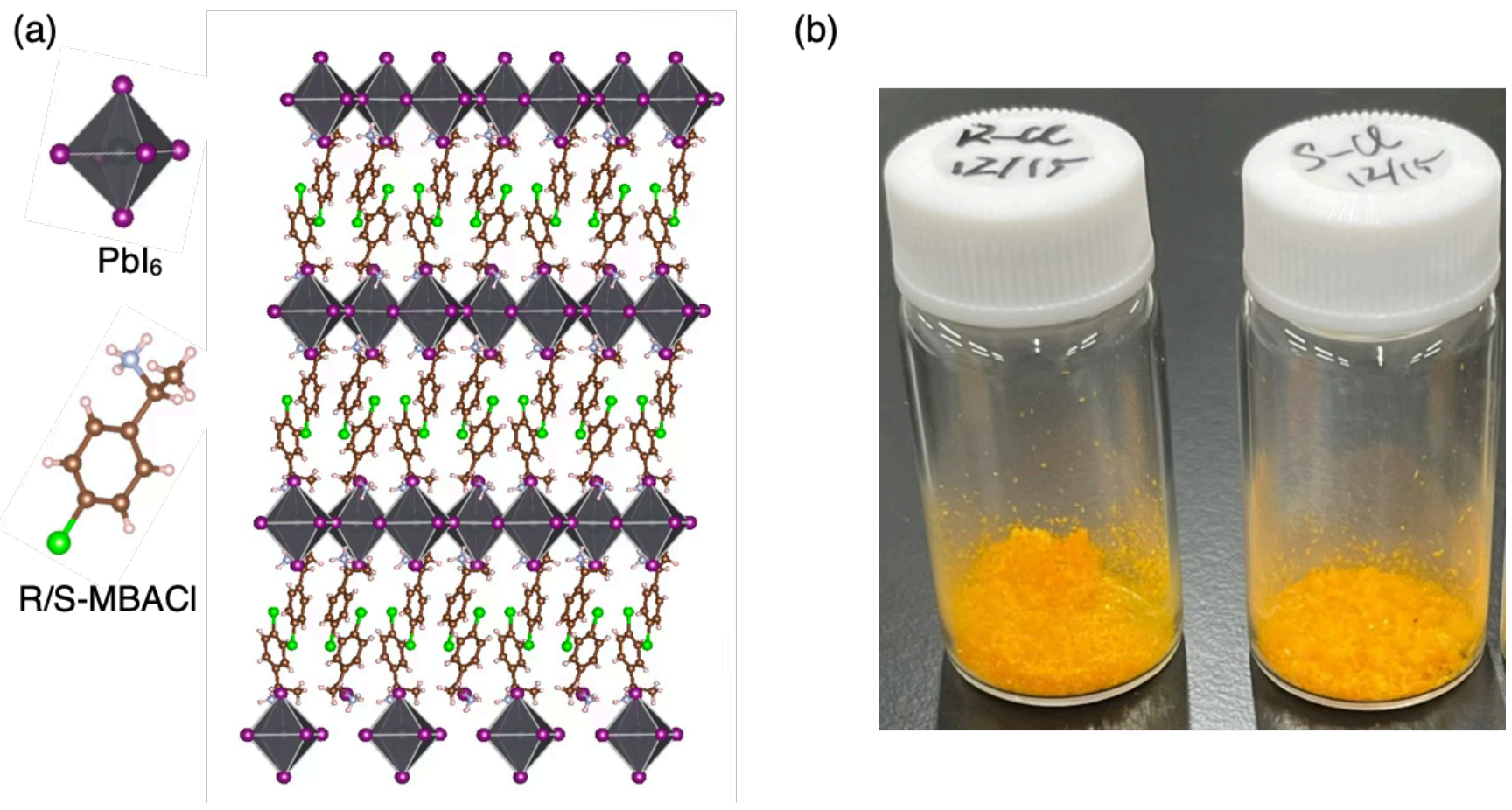

**Figure S1.** (a) Crystalline structure of (R/S-MBACl)$_2$PbI$_4$ single crystal, which is reported in reference [41]. (b) Photograph of synthesized (R/S-MBACl)$_2$PbI$_4$ crystals.

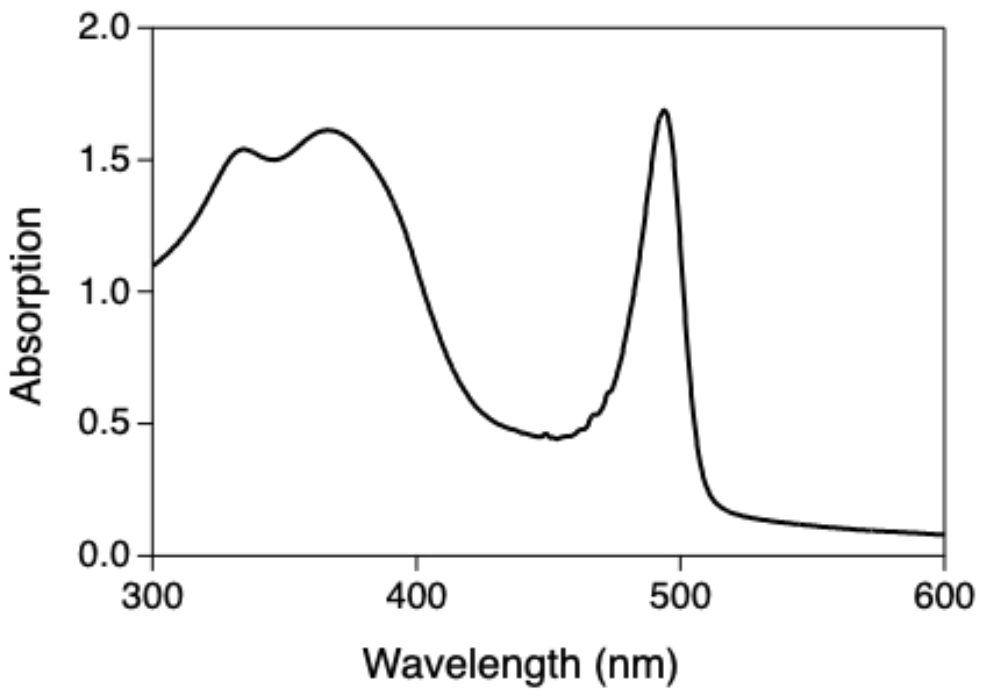

**Figure S2.** Absorption spectrum of R-thin film.

## 2. Optical setup for THG measurement

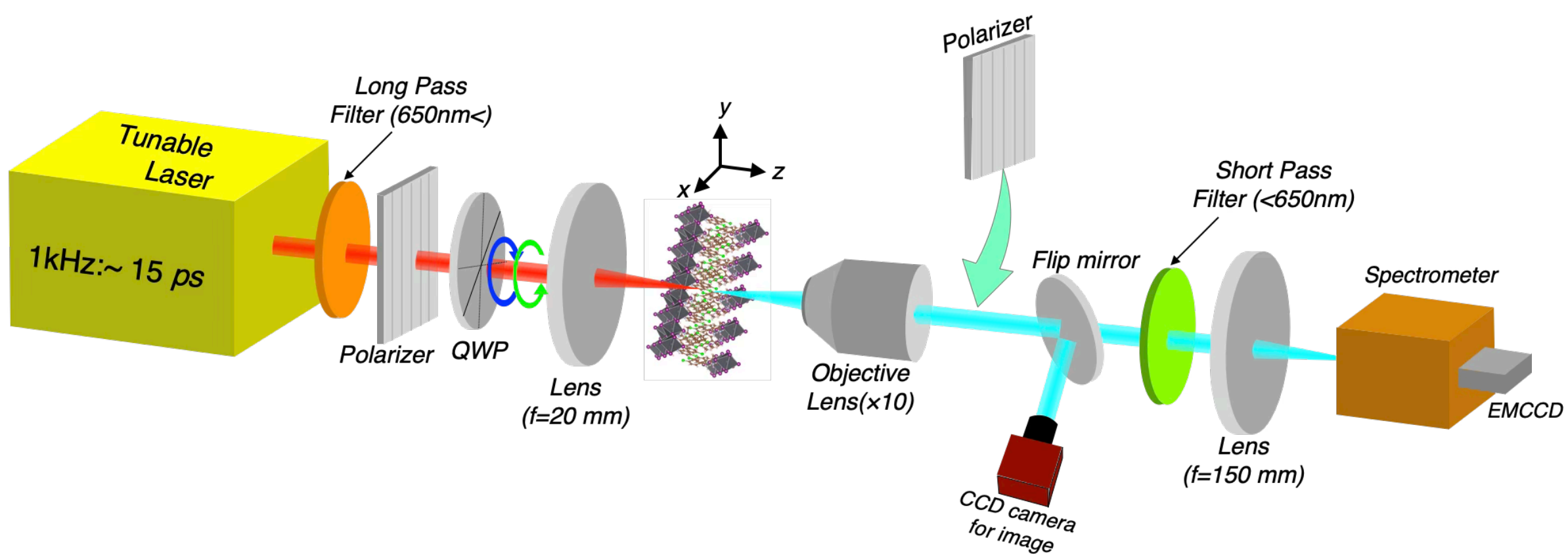

**Figure S3.** Schematic representation of optical set up. Here, *z*-direction is defined as laser propagation direction.

### 3. Comparison with $MoS_2$

In order to compare the effective third-order NLO susceptibility between $(R\text{-}MBACl)_2PbI_4$ and $MoS_2$, the following equation is used. [29]

$$\frac{\chi^{(3)}_{R-SC}}{\chi^{(3)}_{MoS2}} = \left[\frac{(I_3 n_3 n_1{}^3/\omega_3{}^2)_{R-SC}}{(I_3 n_3 n_1{}^3/\omega_3{}^2)_{MoS2}}\left(\frac{\left(\Delta k_{(3)}{}^2 + a_3{}^2/4\right)e^{a_3 L}}{e^{-a_3 L} - 2e^{-a_3 L/2}\cos\left(\Delta k_{(3)}L\right) + 1}\right)_{R-SC}\left(\frac{e^{-a_3 L} - 2e^{-a_3 L/2}\cos\left(\Delta k_{(3)}L\right) + 1}{\left(\Delta k_{(3)}{}^2 + a_3{}^2/4\right)e^{a_3 L}}\right)_{MoS2}\right]^{\frac{1}{2}}$$

where $I_3$ is the THG intensity; $n_3$ and $n_1$ are the refractive index at the THG and fundamental wavelengths, respectively; $\omega_3$ is the angular frequency of the THG; $\alpha_3$ is the absorption coefficient at the THG wavelength; and $\Delta k_{(3)}$ is the wavevector mismatch between the THG wave and the fundamental wave. The refractive index of the 2D perovskite and the refractive index and absorption coefficient of $MoS_2$ were taken from the literature. [44-46]

The effective third-order NLO susceptibility of $MoS_2$ has been reported $(1.7 \pm 0.6) \times 10^{-28}\, m^3/V^2$ at 1560 nm.[45] Thus, at first we estimated NLO susceptibility of $MoS_2$ at 1500 nm and compared it with perovskite thin film.

## 4. THG analysis

### THG-CD analysis for single crystal

The third-order nonlinear optical susceptibility tensor can be expressed in matrix form by allowing permutations of the Cartesian coordinates $x$, $y$, and $z$.

$$\chi^{(3)} = \begin{pmatrix} \chi_{11} & \chi_{12} & \chi_{13} & \chi_{14} & \chi_{15} & \chi_{16} & \chi_{17} & \chi_{18} & \chi_{19} & \chi_{1,10} \\ \chi_{21} & \chi_{22} & \chi_{23} & \chi_{24} & \chi_{25} & \chi_{26} & \chi_{27} & \chi_{28} & \chi_{29} & \chi_{2,10} \\ \chi_{31} & \chi_{32} & \chi_{33} & \chi_{34} & \chi_{35} & \chi_{36} & \chi_{37} & \chi_{38} & \chi_{39} & \chi_{3,10} \end{pmatrix}$$

$$\begin{matrix} xxx & yyy & zzz & yzz & yyz & xzz & xxz & xyy & xxy & xyz \\ 1 & 2 & 3 & 4 & 5 & 6 & 7 & 8 & 9 & 10 \end{matrix}$$

$(R/S\text{-}MBACl)_2PbI_4$ single crystals belong to triclinic $P1$ space group. Owing to the low crystal symmetry, all third-order nonlinear susceptibility tensor components are allowed and independent.

In this study, the $z$-axis is defined as the laser propagation direction. Under normal incidence and normal detection, the tensor components associated with the $z$-direction vanish, and the susceptibility tensor can be simplified as follows:

$$\chi^{(3)} = \begin{pmatrix} \chi_{11} & \chi_{12} & 0 & 0 & 0 & 0 & 0 & \chi_{18} & \chi_{19} & 0 \\ \chi_{21} & \chi_{22} & 0 & 0 & 0 & 0 & 0 & \chi_{28} & \chi_{29} & 0 \\ 0 & 0 & 0 & 0 & 0 & 0 & 0 & 0 & 0 & 0 \end{pmatrix}$$

The third-order nonlinear polarization is described as follows;

$$\begin{pmatrix} P_X^{(3\omega)} \\ P_Y^{(3\omega)} \\ P_Z^{(3\omega)} \end{pmatrix} = \frac{\varepsilon_0}{4} \begin{pmatrix} \chi_{11} & \chi_{12} & 0 & 0 & 0 & 0 & 0 & \chi_{18} & \chi_{19} & 0 \\ \chi_{21} & \chi_{22} & 0 & 0 & 0 & 0 & 0 & \chi_{28} & \chi_{29} & 0 \\ 0 & 0 & 0 & 0 & 0 & 0 & 0 & 0 & 0 & 0 \end{pmatrix} \begin{pmatrix} E_X^3 \\ E_Y^3 \\ E_Z^3 \\ 3E_Y E_Z^2 \\ 3E_Y^2 E_Z \\ 3E_X E_Z^2 \\ 3E_X^2 E_Z \\ 3E_X E_Y^2 \\ 3E_X^2 E_Y \\ 6E_X E_Y E_Z \end{pmatrix}$$

$$= \frac{\varepsilon_0}{4} \begin{pmatrix} \chi_{11} & \chi_{12} & \chi_{18} & \chi_{19} \\ \chi_{21} & \chi_{22} & \chi_{28} & \chi_{29} \end{pmatrix} \begin{pmatrix} E_X^3 \\ E_Y^3 \\ 3E_X E_Y^2 \\ 3E_X^2 E_Y \end{pmatrix}$$

Thus, the THG intensities corresponding to the *x*- and *y*-polarized components are expressed as follows, respectively:

$$I(3\omega) \propto \left[\frac{\epsilon_0}{4}(\chi_{11}E_X^3 + \chi_{12}E_Y^3 + 3\chi_{18}E_X E_Y^2 + 3\chi_{19}E_X^2 E_Y)\right]^2 \text{ (}x\text{-polarization)}$$

$$I(3\omega) \propto \left[\frac{\epsilon_0}{4}(\chi_{21}E_X^3 + \chi_{22}E_Y^3 + 3\chi_{28}E_X E_Y^2 + 3\chi_{29}E_X^2 E_Y)\right]^2 \text{ (}y\text{-polarization)}$$

The total THG intensity is given by the sum of both polarization components.

THG-CD is defined as the intensity difference under fundamental circularly polarized irradiation. The circular polarization states are expressed as

$$E_y = \pm i E_X$$

First, it is clear that the electric field components responsible for the asymmetric THG-CD response under circularly polarized excitation are $E_y^3$ and $E_x^2 E_y$. Accordingly, the existence of the associated tensor components is required for the emergence of THG-CD. For the triclinic $P1$ structure, these tensor components correspond to $\chi_{12}$, $\chi_{19}$, $\chi_{22}$, and $\chi_{29}$.

Substituting $E_y = iE_x$ and $E_y = -iE_x$ into the THG intensity expressions and taking their difference, $\Delta I(3\omega)_{THG-CD}$, yields the THG-CD response:

$\Delta I(3\omega)_x \propto 4Im[(\chi_{11} - 3\chi_{18})(\chi_{12} - 3\chi_{19})^*]$ (x-polarization)

$\Delta I(3\omega)_y \propto 4Im[(\chi_{21} - 3\chi_{28})(\chi_{22} - 3\chi_{29})^*]$ (y-polarization)

The phase difference among the nonlinear susceptibility tensor components gives rise to the anisotropic THG-CD response. Although the triclinic $P1$ structure allows all tensor components to be independent, making it difficult to quantitatively evaluate the contribution of each tensor component individually, the sign reversal of the *g*-factor between the *x*- and *y*-polarized components in $(R/S\text{-}MBACl)_2PbI_4$ single crystals clearly indicates an inversion of the phase relationship among the tensor components contributing to the THG-CD response. This behavior suggests that the tensor components governing the *x*- and *y*-polarized THG responses perceive the structural crystal chirality differently, resulting in opposite chiral NLO responses in the two polarization components.

In this study, the *g*-factor is defined as $g_{THG-CD} = 2\,(I_L - I_R)/(I_L + I_R)$, where $I_L$ and $I_R$ refers to the to the output THG intensity for the fundamental L- or R-CPL, respectively. By expressing $(\chi_{11} - 3\chi_{18}) = Ae^{i\Phi_A}$ and $(\chi_{12} - 3\chi_{19}) = Be^{i\Phi_B}$, the g-factor is expressed as follow;

$$g_{x\ or\ y,THG-CD} = \frac{4|A||B|\sin(\phi_A - \phi_B)}{|A|^2 + |B|^2}$$

The above equation indicates that a giant THG-CD response ($|g_{x\ or\ y,THG-CD} \sim 2|$) can be realized when the symmetric and asymmetric tensor contributions possess comparable amplitudes and a relative phase difference close to $\pm\frac{\pi}{2}$. Specifically, the interference term $Im[(\chi_{11} - 3\chi_{18})(\chi_{12} - 3\chi_{19})^*]$ becomes maximal when $(\phi_A - \phi_B) = \pm\frac{\pi}{2}$, while the additional condition $|A| = |B|$ is required to achieve the theoretical limit of the *g*-factor. Therefore, the experimentally observed giant *g*-factor implies an almost ideal interference condition between the contributing tensor components.

Such a near-ideal interference condition is also evidenced by the polarization conversion from incident CPL to orthogonally polarized THG, because $(\chi_{11} - 3\chi_{18}) = i(\chi_{12} - 3\chi_{19})$ or $(\chi_{21} - 3\chi_{28}) = i(\chi_{22} - 3\chi_{29})$ conditions lead to the near extinction of one THG linear polarization component.

**Fitting for THG intensity as function of QWP rotation angle**

When the rotation angle of the quarter-wave plate (QWP) is defined as $\phi$, the electric field components $E_x$and $E_y$ after the QWP can be expressed using the Jones matrix formalism as

$$\begin{pmatrix} E_x \\ E_y \end{pmatrix} = E_0 \begin{pmatrix} cos^2\varphi + isin^2\varphi \\ \sin\varphi\cos\varphi\,(1-i) \end{pmatrix}$$

The fitting analysis was performed using polarization-resolved THG intensities. As described above, the x-and y-polarized THG intensities are expressed as

$$I(3\omega) \propto \left[\frac{\epsilon_0}{4}(\chi_{11}E_X^3 + \chi_{12}E_Y^3 + 3\chi_{18}E_XE_Y^2 + 3\chi_{19}E_X^2E_Y)\right]^2 \text{ (x-polarization)}$$

$$I(3\omega) \propto \left[\frac{\epsilon_0}{4}(\chi_{21}E_X^3 + \chi_{22}E_Y^3 + 3\chi_{28}E_XE_Y^2 + 3\chi_{29}E_X^2E_Y)\right]^2 \text{ (y-polarization)}$$

Expanding the electric field terms as functions of the QWP rotation angle $\phi$ yields

$$E_x{}^3 = E_0{}^3[(cos^6\varphi - 3cos^2\varphi sin^4\varphi) + i(3cos^4\varphi sin^2\varphi - sin^6\varphi)]$$

$$E_y{}^3 = -2E_0{}^3 sin^3\varphi cos^3\varphi(1+i)$$

$$E_xE_y{}^2 = E_0{}^3[2sin^4\varphi cos^2\varphi - 2isin^2\varphi cos^4\varphi]$$

$$E_x{}^2E_y = E_0{}^3 sin\varphi cos\varphi[(cos^4\varphi - sin^4\varphi + 2cos^2\varphi sin^2\varphi) + i(-cos^4\varphi + sin^4\varphi + 2cos^2\varphi sin^2\varphi)]$$

Here, we consider the nonlinear optical susceptibility tensor to be complex-valued, since the phase differences between the tensor components play an important role in describing the observed anisotropic response. Substituting the complex susceptibility tensor directly into the THG intensity expression results in a lengthy expansion containing numerous combinations of tensor components, making the analytical expression difficult. Therefore, instead of presenting the full expanded form, we express the angular dependence as a Fourier series of harmonic functions. The THG intensity can then be written as

$$I(3\omega) = A + B\cos 2\varphi + C\cos 4\varphi + D\cos 6\varphi + E\cos 8\varphi + F\cos 10\varphi + G\cos 12\varphi \\ + H\sin 2\varphi + I\sin 4\varphi + J\sin 6\varphi + K\sin 8\varphi + L\sin 10\varphi + M\sin 12\varphi$$

where the coefficients A-M are represented by complicated linear combinations of nonlinear susceptibility tensor components.

The fitting value obtained in Figure 3c, d is summarized following table.

| | A | B | C | D | E | F | G | H | I | J | K | L | M |
|---|---|---|---|---|---|---|---|---|---|---|---|---|---|
| **R-x** | 0.462 | -0.023 | 0.406 | -0.044 | 0.043 | 0 | 0 | 0.140 | 0.109 | 0.055 | 0 | 0 | 0 |
| **S-x** | 0.420 | 0 | 0.265 | 0.066 | 0.013 | 0 | 0 | -0.233 | 0.170 | -0.087 | 0.024 | 0 | 0 |
| **R-y** | 0.281 | 0.011 | -0.222 | 0 | -0.012 | 0 | 0 | -0.258 | 0.028 | 0.151 | 0.015 | -0.067 | 0 |
| **S-y** | 0.270 | -0.137 | -0.141 | 0.155 | -0.0468 | -0.033 | 0 | 0.122 | -0.100 | -0.083 | 0.094 | 0.046 | 0 |

**Table S1.** The fitting value obtained in Figure 3c, d

## 5. Linear polarization degree of THG under the CPL irradiation in Figure 3a, b

Here, the degree of linear polarization was defined as stokes parameter, $S_1 = (I_x - I_y)/(I_x + I_y)$.

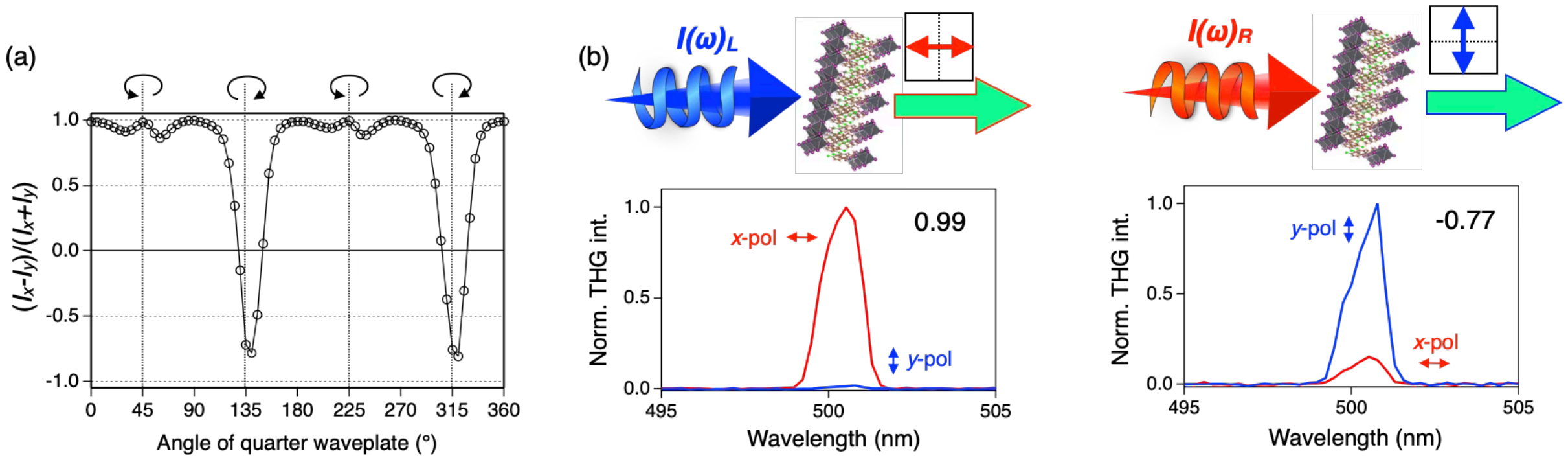

**Figure S4.** (a) Evaluated linear polarization degree as $S_1$ value of stokes parameter as function of QWP angle. (b) Extracted normalized $x$- and $y$-polarized THG spectrum under left and right CPL irradiation.

## 6. Linear polarization degree of THG under the CPL irradiation in Figure 4a, b

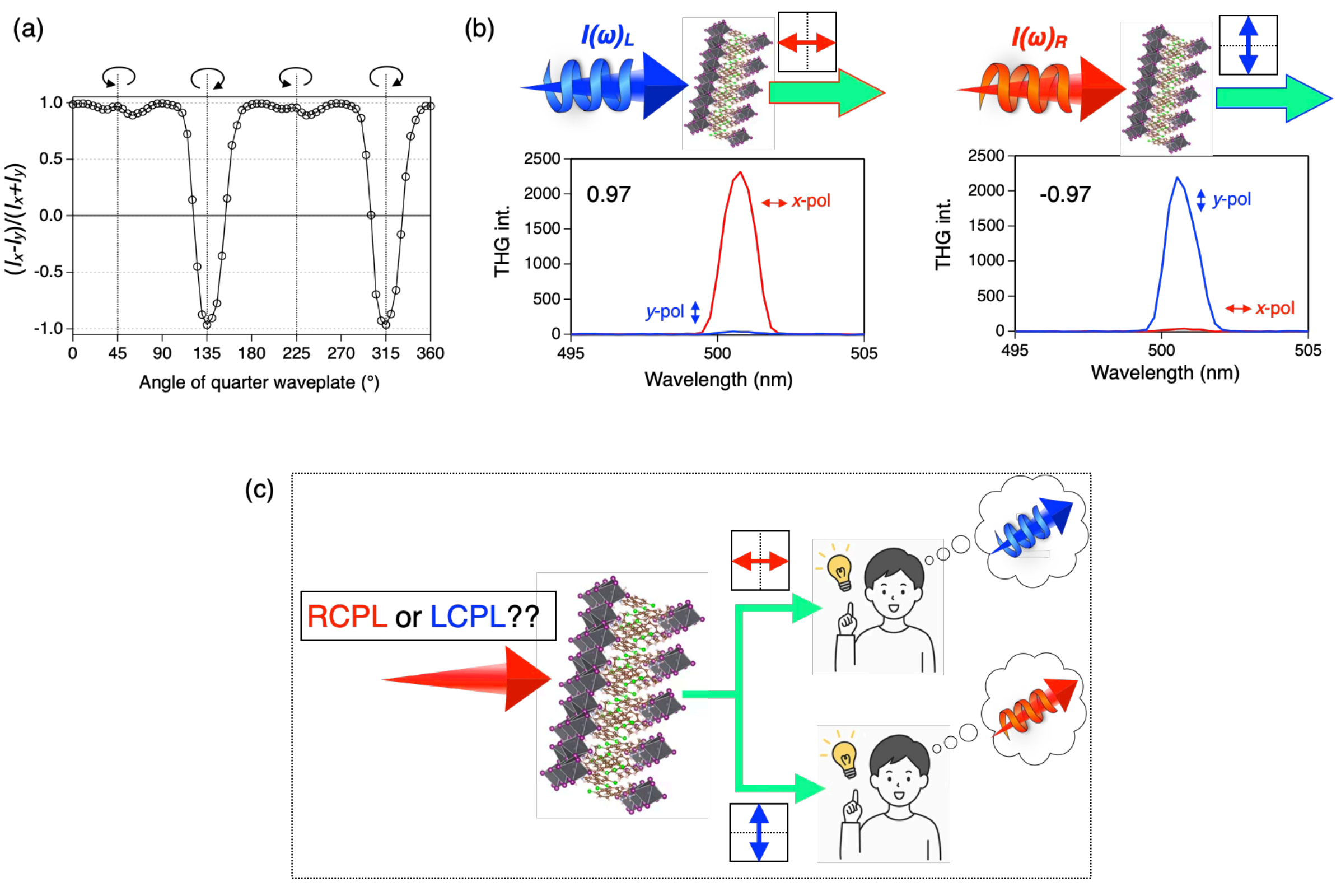

**Figure S5.** (a) Evaluated linear polarization degree as $S_1$ value of stokes parameter as function of QWP angle. (b) Extracted normalized *x*- and *y*-polarized THG spectrum under left and right CPL irradiation. (c) Schematic representation of a system for retrieving the incident circular polarization state from the output linear polarization.

**7. Polarization resolved SHG-CD profile in S-SC sample showing weak $g_{\text{THG-CD}}$.**

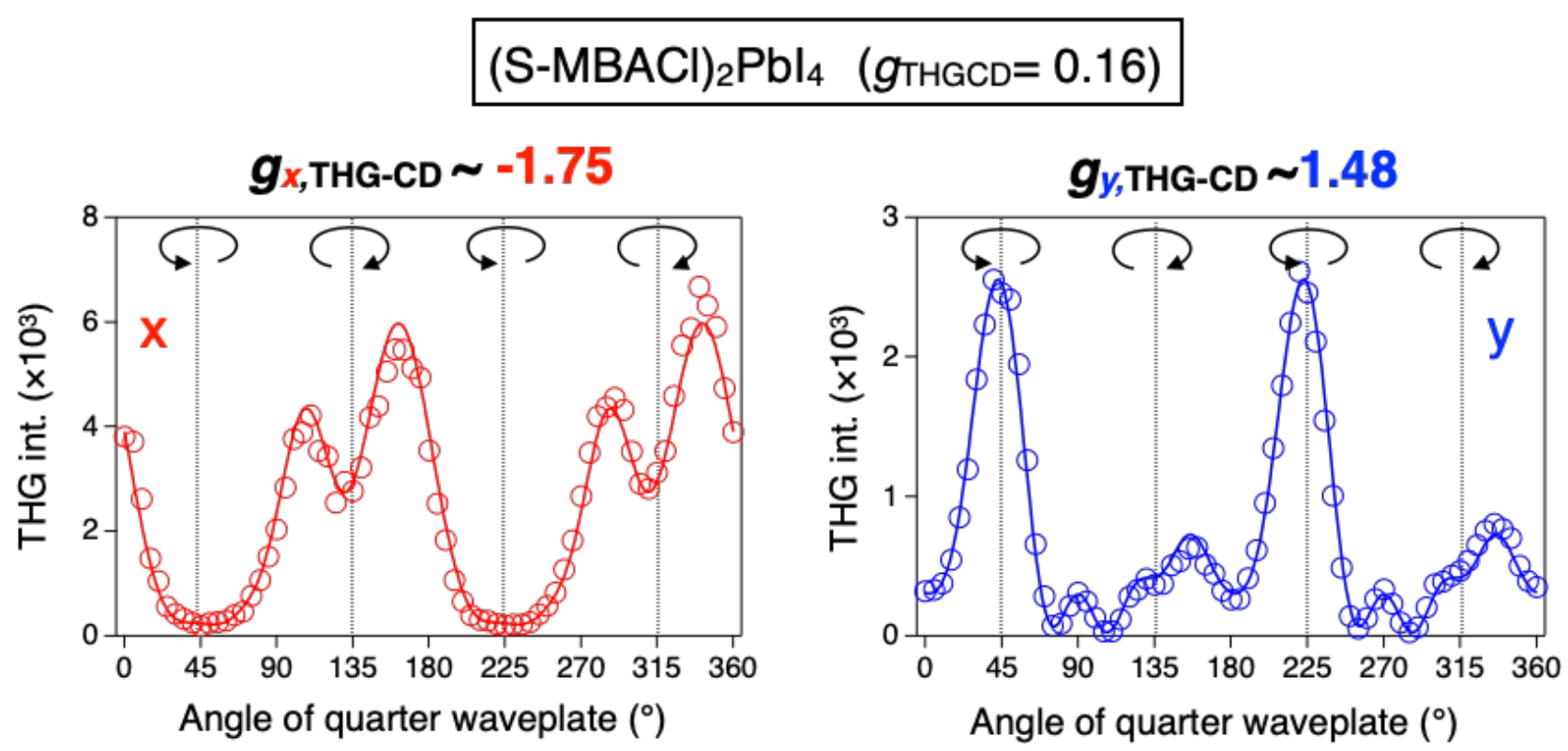

**Figure S6.** Polarization-resolved THG intensity as function of QWP angle in S-SC crystal.

**8. Wavelength dependency of $g_{\text{x,THG-CD}}$ in R- and S-SC sample**

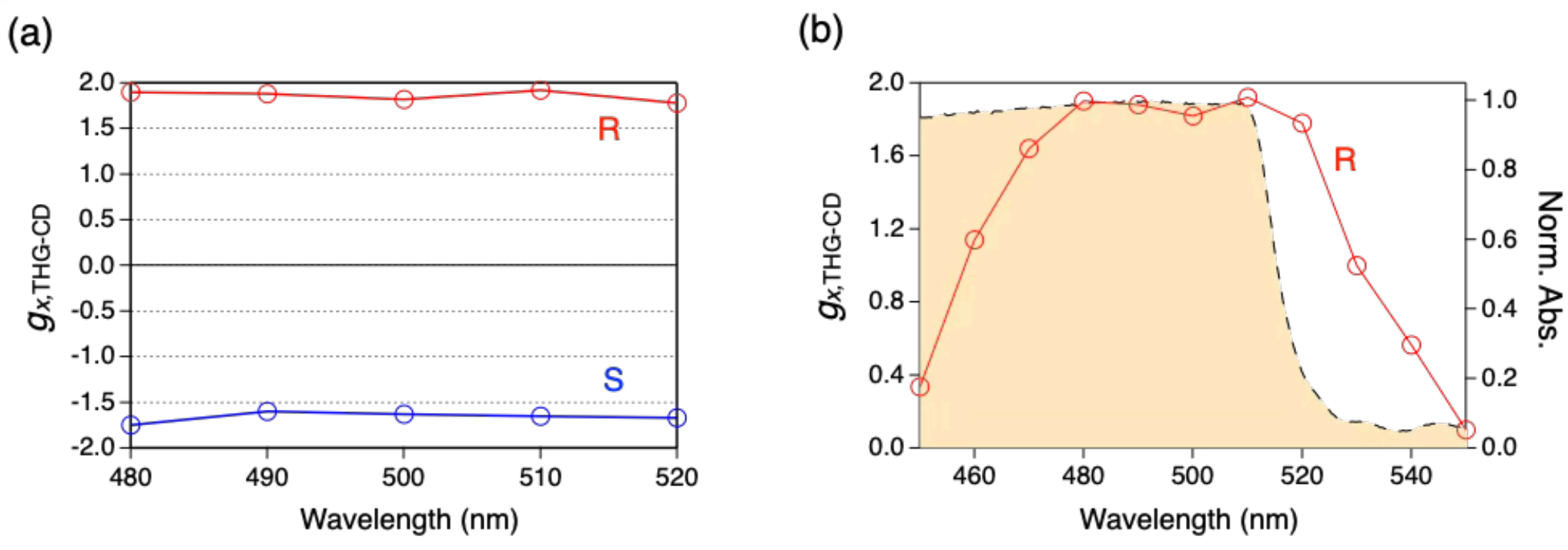

**Figure S7.** (a) Wavelength dependency of $g_{\text{x,THG-CD}}$ in R- and S-SC sample. (b) Wavelength dependency of $g_{\text{x,THG-CD}}$ in R-SC for wider THG wavelength range and normalized absorption spectrum in SC. The large $g_{\text{x,THG-CD}}$ value is only observed near the excitonic resonance wavelength. These measurements were conducted using the same sample as that used for the data presented in Figures 4a, b, and S6.

**9. Chiral THG response in achiral 2D perovskite crystal $(PEA)_2PbI_4$**

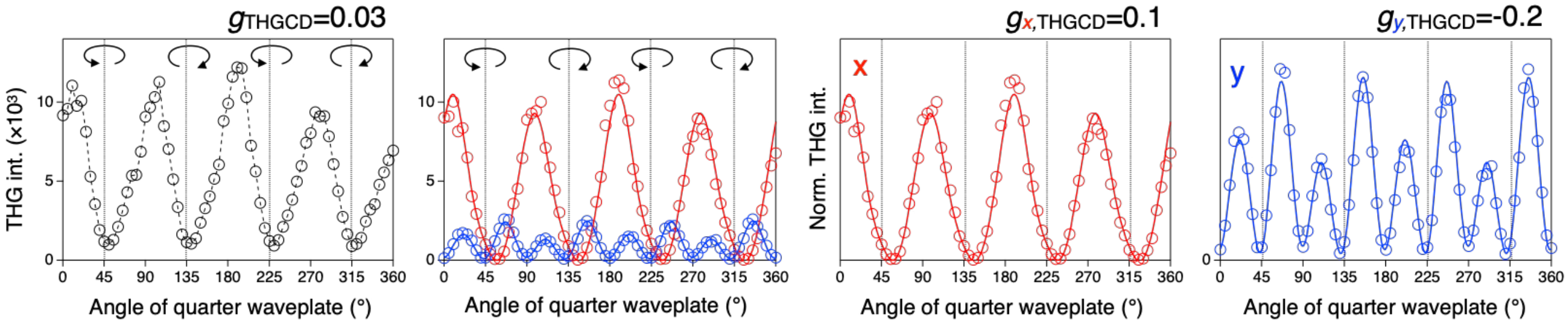

**Figure S8.** Polarization-resolved THG intensity as function of QWP angle in achiral 22D perovskite crystal. The left black dot data show total THG intensity profile and red and blue data show x- and y-polarization data, respectively. The solid line overlapped with x- and y-polarization data is fitting curve obtained using Eq. 4.

**10. THG-CD in $(R/S\text{-}MBACl)_2PbI_4$ thin film**

**THG-CD analysis for thin film**

In the case of thin films, the samples are polycrystalline. Therefore, the crystalline orientation is assumed to be isotropically distributed, and the NLO susceptibility tensor can be approximated by an isotropic form:

$$\begin{pmatrix} P_X^{(3\omega)} \\ P_Y^{(3\omega)} \end{pmatrix} = \frac{\epsilon_0}{4} \begin{pmatrix} \chi_{11} & 0 & \chi_{11} & 0 \\ 0 & \chi_{11} & 0 & \chi_{11} \end{pmatrix} \begin{pmatrix} E_X^3 \\ E_Y^3 \\ E_X E_Y^2 \\ E_X^2 E_Y \end{pmatrix}$$

Considering CPL irradiation ($E_y = \pm i E_X$), both $P_X^{(3\omega)}$and $P_Y^{(3\omega)}$become zero. This indicates that both THG and THG-CD is not expected in the isotropic case under CPL irradiation.

However, as shown in Fig. S9, clear THG-CD responses are experimentally observed in the thin-film samples. Unlike the single crystals, the thin films show the same sign of the chiral response in both polarization channels.

In the poly-crystalline thin-film samples, the small crystal domains are randomly oriented, not perfectly isotropic state, thus we can confirm finite THG activity even under CPL irradiation. In contrast, the polarization-dependent tensor contributions responsible for the sign reversal of the THG-CD response in single crystals are expected to be largely canceled through orientational averaging arising from the random orientation of crystallites. As a result, the THG-CD signal observed in the polycrystalline films is likely to primarily reflect the intrinsic molecular chirality of the material rather than crystallographic symmetry.

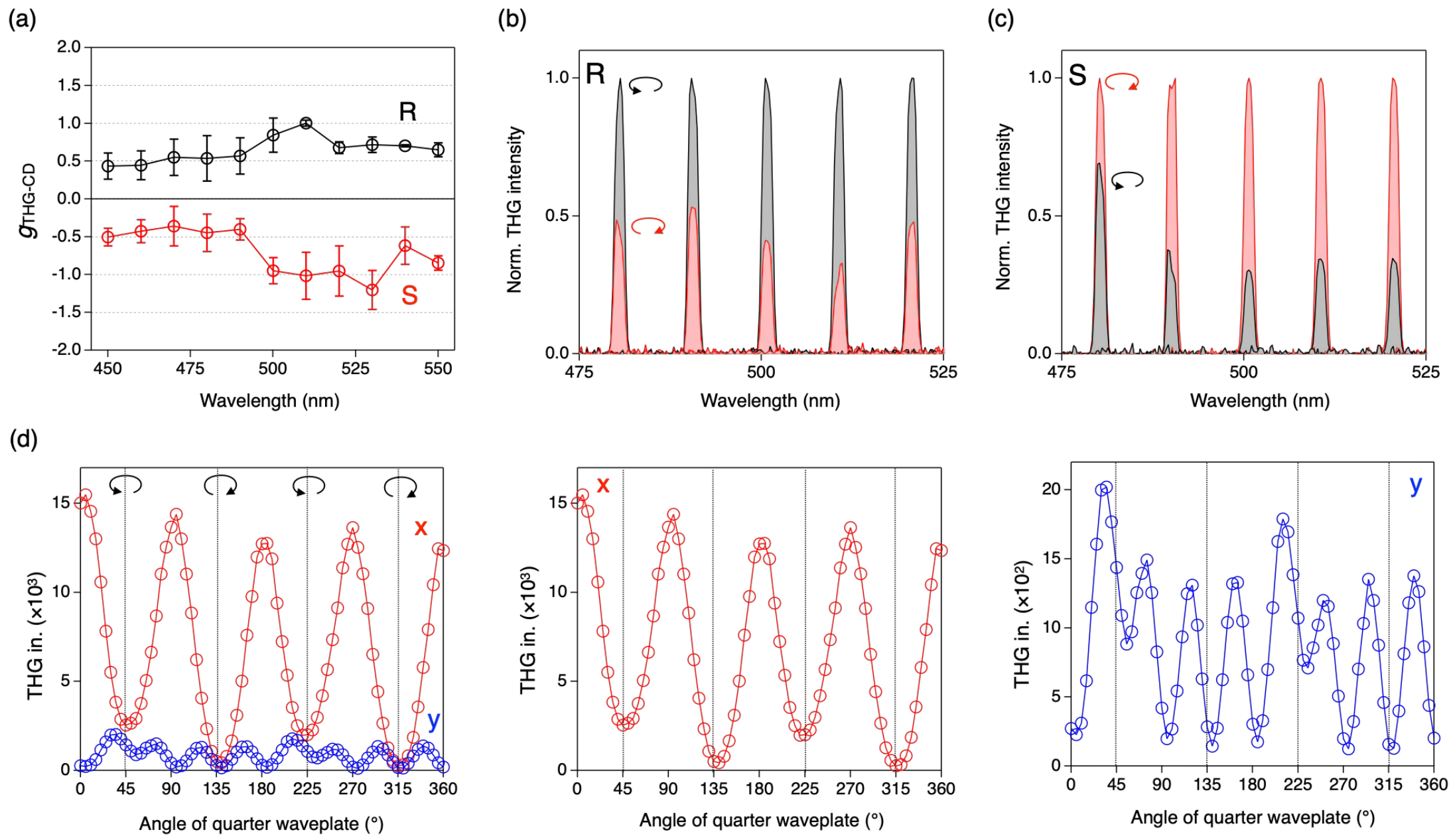

**Figure S9.** (a, b, c) wavelength dependency of g-factor (a) and normalized THG spectrum showing anisotropic response for R- (b) and S-TF samples (c). (d) Polarization resolved THG intensity profiles as function of QWP angles